\documentclass[prb,aps, superscriptaddress,twocolumn]{revtex4-2}
\usepackage{amsmath,amssymb}
\usepackage{graphicx}
\usepackage{wasysym}
\usepackage{amsfonts}
\usepackage{bm}
\usepackage{enumerate}
\usepackage{color}
\usepackage[resetlabels]{multibib}
\usepackage{epstopdf}
\usepackage{latexsym}
\usepackage[breaklinks,colorlinks = true,linkcolor = red,urlcolor=cyan,citecolor=red]{hyperref}
\usepackage[caption=false,singlelinecheck=false]{subfig}
\usepackage[normalem]{ulem}
\usepackage{chemformula}

\usepackage{times}
\newcommand{\bea}{\begin{eqnarray}}
\newcommand{\eea}{\end{eqnarray}}
\newcommand{\be}{\begin{eqnarray}}
\newcommand{\ee}{\end{eqnarray}}
\newcommand{\bw}{\begin{widetext}}
\newcommand{\ew}{\end{widetext}}

\begin{document}
\title{Controllable Skyrmion Islands in a Moir{\'e} Magnet}
\author{Jemin Park}
\email{physjmp@kaist.ac.kr}
\affiliation{Department of Physics, Korea Advanced Institute of Science and Technology, Daejeon, 34141, Korea}
\author{HaRu K. Park}
\email{haru.k.park@kaist.ac.kr}
\affiliation{Department of Physics, Korea Advanced Institute of Science and Technology, Daejeon, 34141, Korea}
\author{SungBin Lee}
\email{sungbin@kaist.ac.kr}
\affiliation{Department of Physics, Korea Advanced Institute of Science and Technology, Daejeon, 34141, Korea}
\date{\today}
\begin{abstract}
Antiferromagnetic(AFM) skyrmions have been in the spotlight as ideal topological magnetic bits. Although they are topologically protected, they do not exhibit the skyrmion Hall effect unlike the ferromagnetic ones. Thus, AFM skyrmions are considered to provide a better control of the skyrmion's motion due to the absence of the skyrmion Magnus effect. In this work, we propose a possible realization of controllable AFM skyrmions in a twisted Moir{\'e} magnet. The tunability of Moir{\'e} materials is not only a good platform for the provision of rich phases, but also for the stabilization of skyrmion phase. We investigate the ground state of twisted bilayer AFM system by solving the Landau-Lifshitz-Gilbert equation in a continuum model. We show that the AFM skyrmions are stabilized even in the absence of the external/dipolar magnetic field, as a consequence of the interplay of interlayer coupling, Dzyaloshinskii-Moriya (DM) interaction and Ising anisotropy. 
More interestingly, due to the magnetoelectric effect, the application of an external electric field locally stabilizes the skyrmions in the twisted bilayer AFM systems, even in the absence of DM interaction. It also allows the skyrmion helicity to change continuously when both the DM interaction and an electric field are present. We show the phase diagram with respect to the strength of interlayer coupling, the DM interaction and an electric field. Our results suggest the possibility of using AFM skyrmions as stable, controllable topological magnetic bits. 
\end{abstract}


\maketitle

\section{Introduction}

Magnetic skyrmions are topologically protected particle-like objects with unique twisted magnetic textures. Due to its topological stability, they have been considered as a promising candidate for topological magnetic bits to store and transmit information in spintronics\cite{Tomasello2014, 10.1063/5.0042917, Kang2016, Legrand2020, https://doi.org/10.1002/adfm.201907140}. They can be stabilized from the ensemble of competing magnetic interactions, such as external/dipolar magnetic field, frustrated exchange interaction and Dzyaloshinnskii-Moriya (DM) interaction.\cite{PhysRevLett.105.197202, PhysRevB.93.064430, ELHOG201832} Their existence has been confirmed not only in the bulk systems but also in thin films, surfaces, or multilayer systems.\cite{Rossler2006, Tong2018, PhysRevB.96.060406} Particularly, antiferromagetic (AFM) skyrmions have been considered as potential information carriers without Magnus force under spin polarized current.\cite{PhysRevApplied.12.064033, PhysRevLett.116.147203, Zhang20161, PhysRevB.100.144439} The absence of the skyrmion Hall effect allows skyrmions to be directly driven by spin-polarized current.\cite{Zhang2016, 10.1063/1.4967006} AFM skyrmions are therefore expected to offer good advantages for overcoming various problems that occur with ferromagnetic skyrmions. Their stabilization and controllability has become the main focus of AFM spintronics.

On another front, the field of twisted van der Waals (vdW) materials has attracted a great deal of attention, providing various exotic phenomena such as superconductivity, Mott insulator and etc.\cite{PhysRevX.13.041049} These phenomena which barely occur on a single layer, are mainly attributed to the interference pattern with enlarged periodic structure. Since such rich phases can be modulated by adjusting the twisting angle, the twisted materials are in the limelight not only because of their rich phases, but also of their controllability\cite{Sivadas2018, PhysRevB.99.144401, SORIANO2019113662}. Particularly, the vdW magnetic materials have been extensively discussed in the context of 2D magnetism and spintronics.\cite{Kim2024, PhysRevB.108.L100401, PhysRevB.104.L100406, doi:10.1073/pnas.2000347117, https://doi.org/10.1002/admi.202300188, PhysRevB.103.L140406} Depending on materials, the vdW materials exhibit not only ferromagnetic but also antiferromagnetic order along either Ising direction or XY plane, and some of them retain their magnetic properties down to a monolayer limit.\cite{Huang2017}.  In multilayer vdW magnets, interlayer exchange interaction reveals rich pattern\cite{doi:10.1073/pnas.2000347117, PhysRevB.104.L100406, Akram2021, kim2022theory, PhysRevResearch.3.013027}, where its coupling constant may change its strength and sign. This opens up a wide variety of magnetic phases that can be created and controlled by the angle of twist between the layers.

In this paper, we suggest the emergence of AFM skyrmions in a twisted bilayer Moir{\'e} magnet. 
First, we take into account interplay of the Moir{\'e} potential and the DM interaction and discuss how the system could stabilize AFM skyrmions. The Moir{\'e} potential with a twist angle gives rise to alternating ferromagnetic and antiferromagnetic inter-layer couplings depending on the region. In the presence of DM interaction, such alternating nature of ferro- and antiferro-region stabilizes the skyrmions only in a particular region. We present the magnetic phase diagram as functions of DM interaction, interlayer coupling and a twist angle. More interestingly, we discuss that in a twisted bilayer Moir{\'e} magnet, the AFM skyrmion can be also stabilized by applying electric field without relying on the DM interaction. This is a consequence of the Moir{\'e} structure and magnetoelectric effect\cite{PhysRevLett.95.057205, Cheong2007, PhysRevB.73.094434}. It allows us to selectively create  AFM skyrmions in a specific region by locally applying electric fields. When both the DM interaction and an electric field are present, the skyrmion helicity is also controllable. Here, the size of AFM skyrmion is controlled by twisting angle. Such AFM skyrmion phenomena are remarkable, with important implications for spintronics and magnetic storage. 
Our work reveals the possibility of the controllable AFM skyrmions in a twisted vdW magetic system, suggesting potential application in AFM spintronics. 

The paper is organized as follows: In section {\ref{sec:2}}, we introduce the continuum Hamiltonian of a twisted bilayer Moir{\'e} magnet. Next, we briefly review the previous studies and illustrate a numerical method, the Landau-Lifshitz-Gilbert equation. In section {\ref{sec:3}}, we show the phase diagram and explain the emergence of the AFM skyrmion phase. We point out unique features of the skyrmion phase and reveal the controllable features of the skyrmion. In section {\ref{sec:4}}, we summarize our result and suggest possible future applications.

\section{Continuum Hamiltonian and Numerical Method}\label{sec:2}

We study a twisted-bilayer of the N{\'e}el type antiferromagnet in a honeycomb lattice. The order parameter of each layer is $\vec{N}_l$, the N{\'e}el vector in layer $l$. Considering slowly varying spin configuration with the N{\'e}el vector, we describe the Hamiltonian in the continuum limit. Taking into account both intra- and inter-layer couplings, the Hamiltonian\cite{doi:10.1073/pnas.2000347117} is written as,

\begin{multline}\label{eq:Original Hamiltonian}\\
H =  \sum_{l=1, 2}[\cfrac{\rho}{2}(\nabla{\vec{N_{l}}})^2-d({N_l^{z}})^2] -J_{\text{Inter}}\Phi(\vec{x})\vec{N_{1}}\cdot\vec{N_{2}} \\
-\sum_{l=1, 2}[J_{\text{DM}}\vec{N_{l}}\cdot(\nabla\times\vec{N_{l}}) + J_{\vec{E}\cdot\vec{P}}N_{l}^z(\nabla \cdot \vec{N_{l}})] .
\end{multline}

Here, $\rho$ is the spin stiffness, which is proportional to the strength of intralayer spin exchange coupling, $d$ is the spin anisotropy perpendicular to the $xy$ plane with $d>0$ and $J_{\text{Inter}}$ is the interlayer exchange coupling, respectively.  $J_{\text{DM}}$ is the intralayer DM interaction and $J_{\vec{E}\cdot\vec{P}}$ is the electric polarization under external electric field.
The DM interaction in the Hamiltonian is generally represented as $\vec{D}_{ij}\cdot(\vec{S}_i\times\vec{S}_j)$, where $\vec{S}_i$ and $\vec{S}_j$ are the spins at site $i$ and $j$ respectively. In this work, we particularly consider the Bloch type DM interaction, represented as $\vec{N_{l}}\cdot(\nabla\times\vec{N_{l}})$ in the continuum limit, with a potential application to transition metal phosphorus trisulphides such as \ch{MnPS3}.\cite{doi:10.1143/JPSJ.52.3919, Wildes_1994, LEFLEM1982455} But we note that other materials with different symmetry may allow the N{\'e}el type DM vector component and similar argument still holds.

We also consider the magnetoelectric effect with $J_{\vec{E}\cdot\vec{P}}$ where the electric polarization is induced by the twisted spin texture in the presence of spin-orbit coupling. The noncollinear spin texture generally gives rise to the electric polarization\cite{PhysRevLett.95.057205, Cheong2007, PhysRevB.73.094434, Tokura_2014}, $\vec{P} \propto \vec{e}_{12} \times (\vec{S_{1}} \times \vec{S_{2}})$, where $\vec{e}_{12}$ indicates a unit vector connecting the two spins, $\vec{S_{1}}$ and $\vec{S_{2}}$. Considering   the external electric field along $\hat{z}$ direction and taking the Taylor expansion of the spin vector up to a leading order, the electric dipole interaction term is represented as $J_{\vec{E}\cdot\vec{P}}N_z(\nabla \cdot \vec{N})$ with the coefficient $J_{\vec{E}\cdot\vec{P}}$ proportional to $E_z$.

Before discussing the DM interaction effect and the magnetoelectric effect, we briefly review Ref \onlinecite{doi:10.1073/pnas.2000347117}, studied in the limit of $J_{\text{DM}} = J_{\vec{E}\cdot\vec{P}} =0$. We introduce the displacement field for $l$-th layer, $\vec{u}_l(\vec{r}) \equiv \vec{r}-\vec{r}_0$ where $\vec{r}_0$ is the original lattice position before twist and $\vec{r}$ is the deformed position after twist. Then one can write the spin density in terms of the leading order of the Fourier series, $\vec{S}_l(\vec{r})=n_0\vec{N}_l\sum_{l}\sin(\vec{b}_i\cdot (\vec{r}-\vec{u}_l))$, where $\vec{b}_{i}$ are the three reciprocal vectors of honeycomb lattice and  $n_0$ is the magnitude of the ordered moment. When the layers are twisted with small angle $\theta$, we get $\vec{u}_1(\vec{r})=-\vec{u}_2(\vec{r})=\frac{\theta}{2}\hat{z}\times \vec{r}$. Now inserting these expressions into the interlayer coupling Hamiltonian $\vec{S}_1(\vec{r})\cdot \vec{S}_2(\vec{r})$, keeping the long wavevector terms only, gives $J_{\text{Inter}}$-term in Eq. \ref{eq:Original Hamiltonian} with,
\begin{equation}
\Phi(\vec{x}) = \sum_{i=1}^3 \cos(\vec{q}_i \cdot \vec{r}),
\end{equation}
where $\vec{q}_i=\theta \hat{z}\times \vec{b}_i$ are the Moir{\'e} lattice reciprocal vectors.
The sign of the $\Phi(\vec{x})$ determines ferromagnetic or antiferromagnetic interlayer coupling of $\vec{N_l}$. This interlayer potential represents the periodicity of the Moir{\'e} pattern.

\begin{figure*}
\includegraphics[width=1\textwidth]{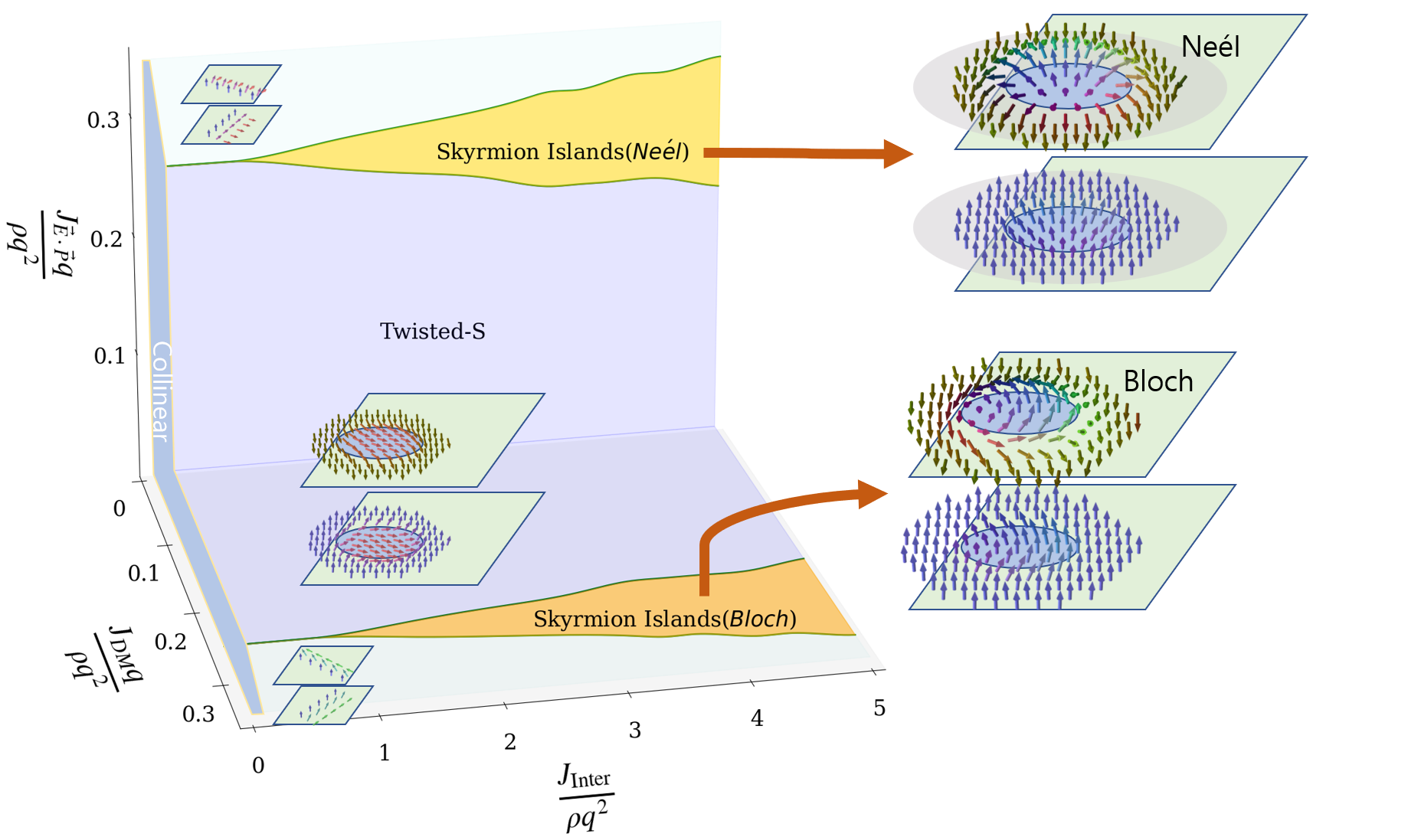}
\caption{Phase diagram as functions of the dimensionless parameters $J_{Inter}/({\rho} q^2)$ and $J_{\text{DM}}q/({\rho} q^2)$ where $d/({\rho} q^2)=3/40$ is chosen. Depending on the relative strength between interlayer coupling, electric dipole and DM interaction, distinct phases are stabilized; Collinear, Twisted-S, Skyrmion Islands. For very small interlayer coupling, the collinear phase is stabilized where the N{\'e}el vectors are simply aligned along $\pm \hat{z}$ in each layer.
When interlayer coupling becomes large but electric dipole and DM interactions are small, the Twisted-S phase is stabilized. In this phase, the N{\'e}el vectors are aligned in the $xy$ plane inside each island, while they are aligned along the $z$ axis outside the islands.
With larger dipole/DM interaction, the interplay of the Moir{\'e} potential and the DM interaction or electric dipole interaction becomes significant.
In this regime, the AFM skyrmions are stabilized within the islands in a single layer. For finite DM interaction, the Bloch type AFM skyrmions are stabilized, while for finite electric dipole interaction, the N{\'e}el type AFM skyrmions are stabilized.
In the N{\'e}el type AFM skyrmion phase, the skyrmions are locally stabilized only in the region where an electric field is applied.
The gray circle in the N{\'e}el type AFM skyrmion phase represents the external electric field applied out of the plane.
In large electric-dipole interaction or DM interaction limit, they ignore the Moir{\'e} potential and the N{\'e}el vectors form incommensurate structure beyond the islands. Thus, there is no Moir{\'e} structure as in the former case.
The image in each phase is the simulation result, describing the configuration of the N{\'e}el vectors. The rounded blue region in each image represents the island where the N{\'e}el vectors between the layers interact ferromagnetically.}
\label{fig:phase diagram}
\end{figure*}

In the absence of the external electric field and the DM interaction, there are three magnetic orders: collinear phase, twisted-A phase and twisted-S phase. These orders depend on the relative magnitude of the coefficients $\rho, d,$ and $J_{\text{Inter}}$. In a small interlayer coupling regime, a collinear phase is stabilized. In this case, $J_{Inter}$ term cannot overcome the energy from the spin stiffness $\rho$. With the anisotropy $d\neq0$, the $\vec{N}_l$ are simply aligned along $+\hat{z}$ or $-\hat{z}$ in each layer. Thus, there is no Moir{\'e} structure. On the other hand, when the interlayer coupling $J_{\textrm{Inter}}$ becomes stronger, it overcomes the penalty of the stiffness.
Since $J_{\textrm{Inter}}$-term includes $\Phi(\vec{x})$, the Moir{\'e} structure comes in. The feature of $\Phi(\vec{x})$ whose periodicity depends on the Moir{\'e} lattice reciprocal vectors gives enlarged unit cell, and the sign of $\Phi(x)$ determines the region where ferromagnetic or antiferromagnetic interlayer coupling dominates. We will refer to the region where the interlayer coupling is ferromagnetic as the \textit{island}. In this large $J_{\textrm{Inter}}$ limit, there are two different magnetic orders.
For large $d$ limit, the Twisted-A phase is stabilized, where the N{\'e}el vectors are ferromagnetically or antiferromagnetically aligned along the $\pm \hat{z}$ inside or outside the islands. 
For small $d$ case, the interplay of stiffness and anisotropy stabilizes the Twisted-S phase, where the N{\'e}el vectors inside the islands are ferromagnetically aligned along the $xy$ plane, but the N{\'e}el vectors outside the islands are antiferromagnetically aligned along the $\pm \hat{z}$ axis between the layers. In Fig.\ref{fig:phase diagram}, the Twisted-S phase is illustrated in the purple region.

Having understood the magnetic ordering in the limit of $J_{\text{DM}} = J_{\vec{E} \cdot \vec{P}}=0$, now let's discuss the magnetic phases when the DM interaction and the magnetoelectric effect are present.
In order to discuss the magnetic ordering, we adopt the Landau-Lifshitz-Gilbert (LLG) equation to minimize the energy functional of Eq.\eqref{eq:Original Hamiltonian}, and look for the ground state. The LLG equation is represented as,
\begin{equation}\label{eq:LLG}
\frac{d\vec{N_l}}{dt}=\gamma\vec{N_l}\times\vec{H}_{eff}-\lambda\vec{N_l}\times(\vec{N_l}\times\vec{H}_{eff}),
\end{equation}
where $\vec{H}_{eff}$ is the effective field written as,
\begin{equation}\label{eq:eff}
\vec{H}_{eff}=-\cfrac{\delta H}{\delta\vec{N_l}}.
\end{equation}
Here, $\gamma$ and $\lambda$ are the gyromagnetic ratio and the damping parameter respectively. We adjust damping parameter and relax the various initial state. 
Our numerical system includes 1200 meshes which are made up of three Moir{\'e} unit cells with a hexagon shape. Under the periodic boundary condition, various initial conditions for the N{\'e}el vectors  have been simulated, 
 including random, uniform, twisted-S/A  and Skyrmion configurations. 

\section{chiral magnetic texture in the Moir{\'e} pattern and its controllable properties} \label{sec:3}
The phase diagram is shown in Figure \ref{fig:phase diagram} with respect to $J_{\text{DM}}q/({\rho} q^2)$, $J_{\text{Inter}}/\rho q^2$, and $J_{\vec{E} \cdot \vec{P}}q/({\rho} q^2)$, with fixed value of $d/({\rho} q^2)=3/40$. Here, $q=|\vec{q}_i|$, the magnitude of the Moir{\'e} reciprocal vectors, $\vec{q}_i$. We note that the phase diagram does not qualitatively change under the small change of $d/({\rho} q^2)$. 

For a small value of $J_{\text{Inter}}$ in the limit of $J_{\vec{E} \cdot \vec{P}}=J_{\text{DM}}=0$, a typical N{\'e}el type collinear ordering emerges in each layer, as shown in Fig.\ref{fig:phase diagram}, also discussed in the previous section.
On the other hand, for larger $J_{\text{Inter}}$, the Moir{\'e} structure becomes important and stabilizes the Twisted phase as discussed in Section \ref{sec:2}. Particularly with a given magnitude of spin anisotropy $d$ we set here, the Twisted-S phase is stabilized having magnetic ordering in the $xy$ plane within the islands region (for FM Moir{\'e} region) and the Ising ($\pm\hat{z}$-axis) ordering outside the islands (for AFM Moir{\'e} region).

Now let's discuss the case in which the DM interaction or the magnetoelectric interaction are finite. 
In the following, we first argue the case of $J_{\text{DM}} \neq 0$ and $J_{\vec{E} \cdot \vec{P}}=0$, and then discuss the case of $J_{\text{DM}} = 0$ and $J_{\vec{E} \cdot \vec{P}}\neq0$. Finally, we argue the case when both $J_{\text{DM}}$ and $J_{\vec{E} \cdot \vec{P}}$ are nonzero.

\begin{figure}[h]
\includegraphics[width=0.4\textwidth]{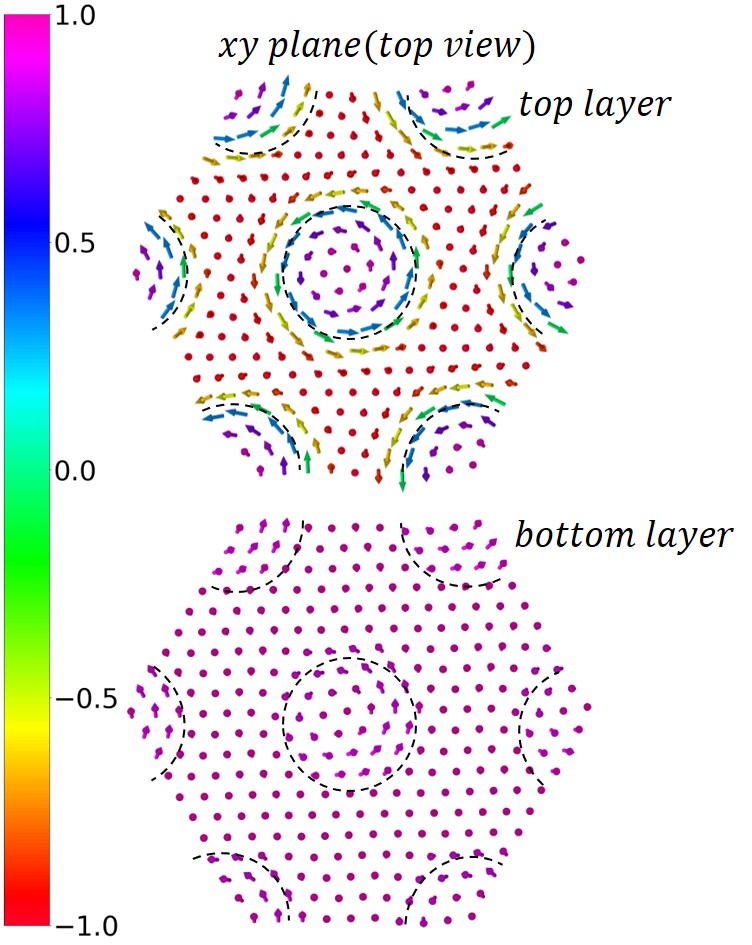}}
\caption {Configuration of the N{\'e}el vector, $\vec{N}_l$, for the top and bottom layers in the Bloch type skyrmion island phase when $J_{\text{DM}}q/({\rho} q^2)=0.27$ and $J_{\vec{E}\cdot \vec{P}}=0 $. The numerical simulation is performed with 1200 lattice points and the figures are thumbnails of the simulation. Arrows and colors indicate the direction of $\vec{N}_l$  in the plane and out of the plane respectively. The Bloch type AFM skyrmion is stabilized only in the top layer, while  the N{\'e}el vectors are slightly tilted from $+\hat{z}$, giving an umbrella like configuration in the bottom layer. The helicity of the skyrmion, $\eta$, is $\pi/2$. Dashed lines are for the boundary of the AFM skyrmion islands region. The color bar indicates the magnitude of $z$ component (out of plane) of the N{\'e}el vectors. }
{\label{fig:orderingDM}
\end{figure}

Now we first consider how the DM interaction changes the magnetic structure. For very strong DM interaction limit, it overcomes the Moir{\'e} potential and  the non-coplanar N{\'e}el vectors are stabilized over the islands, which is beyond our interest. For intermediate values of the DM interaction, the interplay of the Moir{\'e} potential and the anisotropic spin interaction hosts a chiral magnetic structure and gives topological spin texture, the skyrmion.
As the Moir{\'e} potential makes the boundary between ferromagnetically and antiferromagnetically interacting regions and forms the island, a magnetic domain wall is also stabilized along the boundary. Along this domain wall, the AFM skyrmion structure is stabilized, while outside this domain wall, the N{\'e}el vectors are simply aligned along $\pm \hat{z}$. Since we have considered the Bloch type $J_{\text{DM}}$, the Bloch-type AFM skyrmion is stabilized within the island, shown as the Bloch type AFM skyrmion island phase in Fig.\ref{fig:phase diagram}.
In detail, the AFM skyrmions are described as following.
Consider the polar coordinate $(r,\varphi)$ whose origin at the center of the Skyrmion. When we represent the Skyrmion configuration as $\vec{N}(\vec{r})=(\sin{\theta}\cos{\phi}, \sin{\theta}\sin{\phi}, \cos{\theta})$ with polar angle $\theta$ and azimuthal angle $\phi$, the angle $\theta$ only depends on $r$, while the angle $\phi$ can be written in the form as $\phi=n\varphi+\eta$, where integer $n$ is the vorticity and $\eta$ is the helicity.\cite{Zhang_2020} The skyrmion configuration described by $\eta=0$ or $\pi$ is the N{\'e}el type, whereas, the a skyrmion with $\eta=\pi/2$ or $\eta=-\pi/2$ is the Bloch type.

Interestingly, in this phase, the AFM skyrmions confined within the islands are stabilized only in a single layer. Whereas, in another layer, the order parameter is almost aligned along $\hat{z}$ axis. Figure \ref{fig:orderingDM} presents the top view of the magnetic structure in each layer.
Asymmetric skyrmion structure between the layers can be understood as the difference of the interaction inside and outside of the islands. Outside the islands, due to the interplay of antiferromagnetic interlayer coupling and anisotropy, the N{\'e}el vectors, $\vec{N_l}$, should be anti-aligned along $\hat{z}$ and $-\hat{z}$ in each layer. Inside the islands, if the skyrmions are stabilized in both layers, the configuration of the skyrmions must be the same due to the ferromagnetic interlayer coupling. Then, at the boundary, the N{\'e}el vectors should point in the same direction at both layers. However, since the N{\'e}el  vectors outside the islands point in the opposite directions due to the antiferromagnetic interaction, it induces a drastic flipping of spins at the boundary of the islands at least in one layer. It gives rise to a big energy penalty, and is therefore not preferred. On the other hand, suppose that only a single layer hosts skyrmion and the order parameter in the other layer aligned along $\hat{z}$ axis. In this case, there is no such drastic flipping of spins at the boundary of the islands, since the order parameter in one layer is pointing $+\hat{z}$ and the one in another layer is pointing $-\hat{z}$. Thus, the  N{\'e}el  vectors can be smoothly connected at the boundary between inside and outside of islands in this case, resulting in low energy cost from the stiffness term. Therefore, the skyrmion configuration only in a single layer is favored.

\begin{figure}[b]
\includegraphics[width=0.45\textwidth]{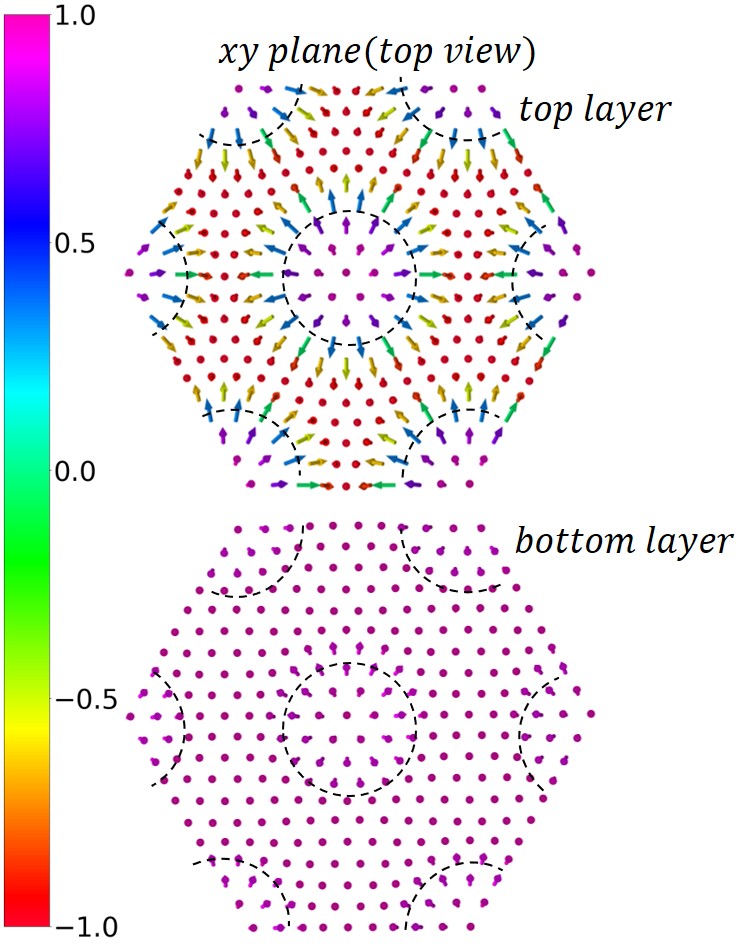}}
\caption {Configuration of the N{\'e}el vector, $\vec{N}_l$, in the top and bottom layers for the N{\'e}el type skyrmion island phase, when  $J_{\vec{E}\cdot \vec{P}}=0.27$ and $J_{\text{DM}}=0$. Similar to Fig.\ref{fig:orderingDM}, arrows and colors indicate the direction of $\vec{N}_l$  in the plane and out of the plane respectively. The N{\'e}el type AFM skyrmion with helicity $\eta=0$, is stabilized only in the top layer, while  the N{\'e}el vectors are slightly tilted from $+\hat{z}$, giving an umbrella like configuration in the bottom layer. Since $J_{\vec{E}\cdot \vec{P}}$ term and $J_{\text{DM}}$ term has the relation of $\pi/2$ spin rotation along $\hat{z}$, the N{\'e}el vectors in this configuration is also equivalent to $\pi/2$ rotation along $\hat{z}$ from the Bloch type skyrmion configuration shown in Fig.\ref{fig:orderingDM}. The color bar indicates the magnitude of $z$ component (out of plane) of the N{\'e}el vectors. }
{\label{fig:EP}
\end{figure}

Another feature of the AFM skyrmion island phase is the controllability of the skyrmion radius. The skyrmions are bounded by the island which is determined by the Moir{\'e} potential. Since the Moir{\'e} potential depends on the Moir{\'e} lattice length scale which is determined by the twisting angle, the skyrmion radius can be controlled by adjusting the twisting angle.

Since we can allocate only one skyrmion among the two layers, the ground state has degeneracy from the choice of the skyrmion in each islands and the swapping layer degrees of freedom. Figure \ref{fig:orderingDM} shows one of such configuration. For example, changing the sign of the N{\'e}el vector in a skyrmion and translating to the bottom layer gives the other ground state. Thus, there exists two energy equivalent configuration per island.

So far, we have discussed the AFM skyrmion phase with a finite $J_{\text{DM}}$ and $J_{\vec{E} \cdot \vec{P}}\!=\!0$. Now let's discuss the case with a finite $J_{\vec{E} \cdot \vec{P}}$ and $J_{\text{DM}}\!=\!0$.  
Unlike the DM interaction, the electric dipole interaction due to the magnetoelectric effect is a controllable parameter by adjusting the strength and domain of the external electric field. Since the coefficient of $J_{\vec{E} \cdot \vec{P}}$ is proportional to the applied electric field, we emphasize that,  depending on where the electric field is applied,  the magnetic phase transitions can be locally manipulated. Fig.\ref{fig:phase diagram} show the  phase transitions from collinear to Twisted-S and to the N{\'e}el type AFM skyrmion island phase depending on the relative values of $J_{\text{Inter}}$ and  $J_{\vec{E} \cdot \vec{P}}$. In Fig.\ref{fig:phase diagram}, the gray area in the image of the N{\'e}el type skyrmion island indicates the area of the applied electric field. Thus, the magetoelectric effect can locally stabilize the N{\'e}el type AFM skyrmions owing to applied electric field. 

In Eq.\eqref{eq:Original Hamiltonian}, we note that the difference between the DM interaction and the electric-dipole interaction comes from the divergence or curl acting on the vector field, and their difference is equivalent to $\pi/2$ spin rotation along $\hat{z}$ axis.
Considering $\pi/2$ spin rotation is performed along $\hat{z}$, $N_l^{x}$ and $N_l^{y}$ are transformed into $-N_l^y$ and $N_l^x$ respectively. Thus, the DM interaction $N_l^{z}(\partial_{x}N_l^{y}-\partial_{y}N_l^{x})$ transforms into $N_l^{z}(\partial_{x}N_l^{x}+\partial_{y}N_l^{y})$ which is the electric dipole interaction.
Hence, the argument for the DM interaction also holds for the electric dipole interaction case by $\pi/2$ rotation. 
For a finite $J_{\vec{E} \cdot \vec{P}} $ and $J_{\text{DM}}=0$ case, the N{\'e}el type skyrmion is stabilized with the helicity $\eta=0$, as shown in  Fig.\ref{fig:EP}. 

\begin{figure}[h]
\includegraphics[width=0.45\textwidth]{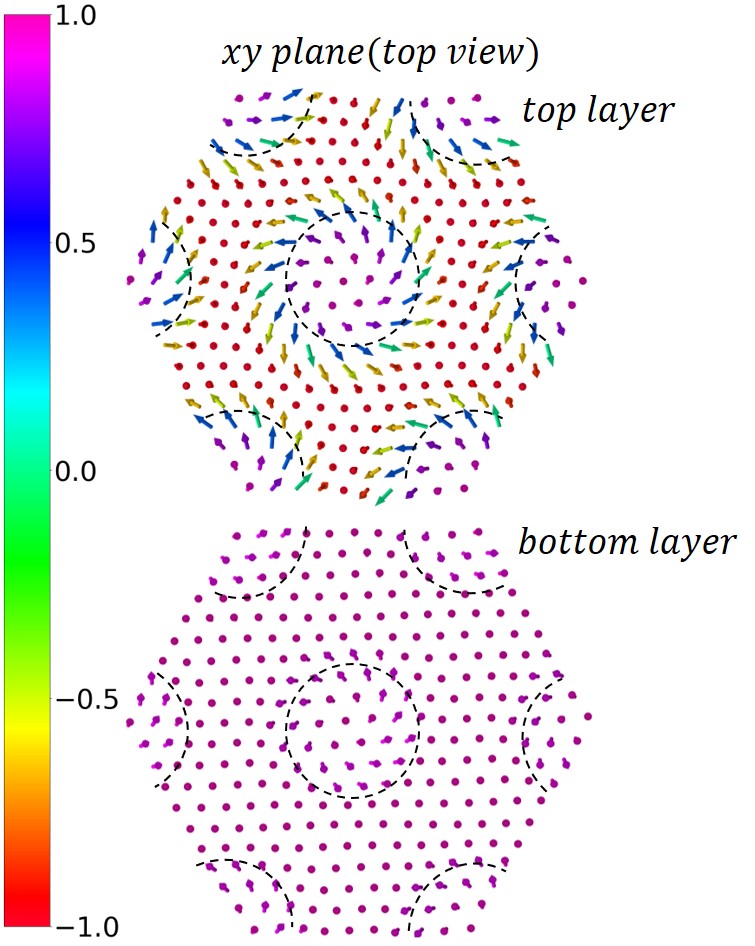}
\caption {Configuration of the N{\'e}el vector, $\vec{N}_l$, in top and bottom layers when $J_{\text{DM}}=J_{diplole}=0.2$. The helicity which stabilizes skyrmion is given by $\pi/4=\arctan(J_{\text{DM}}/J_{\vec{E}\cdot\vec{P}})$. See the main text for details. Unlike the DM interaction, the strength of an electric field is adjustable factor. Thus, this helicity can be controlled via external electric field strength.}
{\label{fig:helicity}}
\end{figure}

Next, we move on to the case when both $J_{\text{DM}}$ and $J_{\vec{E} \cdot \vec{P}}$ are nonzero. In the skyrmion phase, the energy gain from the DM interaction and electric polarization can be written as $E_{0}(J_{\text{DM}}\sin\eta+J_{\vec{E} \cdot \vec{P}}\cos\eta)$ where $\eta$ is the skyrmion helicity.
 Thus under certain magnitudes of $J_{\text{DM}}$ and $J_{\vec{E}\cdot\vec{P}}$, energy summed over DM interaction and electric polarization is minimized when the helicity is $\eta=\arctan(J_{\text{DM}}/J_{\vec{E}\cdot\vec{P}})$. The skyrmion helicity continuously varies under the electric field strength, allowing us to control the skyrmion helicity. Figure \ref{fig:helicity} shows the example of the chiral magnetic structure for the case of  $J_{\text{DM}}=J_{\vec{E}\cdot\vec{P}}$. The helicity of the skyrmion is $\pi/4$, modulated from $\pi/2$ or $0$. 

\section{Conclusion}\label{sec:4}
AFM skyrmions have been considered as an ideal information carrier in spintronics. Contrast to the ferromagnetic ones, they don't exhibit the skyrmion Hall effect and thus provide a better control of their motion. To stabilize such AFM skyrmion, we propose a twisted bilayer magnetic system as a platform to build AFM skyrmions. Taking into account the interplay of intralayer, interlayer coupling, DM interaction and magnetoelectric effect, we discuss the spin model in the continuum limit.  The controllability of the Moir{\'e} system leads to the different competing energy scales between the electric polarization, the DM interaction and the Moir{\'e} potential. Such competition gives rise to a wealth of phases, including the chiral Moir{\'e} structure, the N{\'e}el type or Bloch type skyrmion island phase. The interplay of DM interaction and Moir{\'e} potential stabilizes the Bloch type skyrmions, whereas, the magnetoelectric coupling can stabilize the N{\'e}el type skyrmions. 
 In particular, due to the magetoelectric effect, the N{\'e}el type skyrmions can be locally stabilized confined within an applied electric field. It suggests the use of electric fields to manipulate the AFM skyrmions. 
 
 The ability to locally stabilize and confine the Néel-type skyrmions within an applied electric field is significant for potential applications. It implies that the electric field can be used as a tool to manipulate and control the formation, motion, and annihilation of these skyrmions in a controlled manner.
 Since skyrmion is bounded by Moir{\'e} potential, position and radius of the skyrmion is ruled by a twisting angle, which is a unique feature of the Moir{\'e} magnets. Furthermore, we have shown that by applying an electric field on each skyrmion, certain helicity configuration is favored depending on the relative strength between DM interaction and electric polarization. 
 
 Considering the coupling between such AFM skyrmion island phases and the conducting system, which may allow us to control conductivity, could be an interesting subject for future studies. We expect these controllability of skyrmion phases offer a platform for spintronics applications, including skyrmion-based logic gates, memory devices, and so on.\cite{Tikhonov2020} In addition, it would be interesting to study possible control of skyrmion dynamics by other factors not considered in this paper, e.g., current, external magnetic field and strain. Furthermore, stacking more than two layers allows additional adjustable degrees of freedom, providing a much richer set of potential patterns between layers. It would give possibility to stabilize other interesting magnetic orderings e.g., merons, exotic domain wall structures and perhaps, topologically protected orderings with higher winding number than one. In another aspect, engineering quantum bits based on this system would be a good starting point for further research. Since there is a degree of freedom in choosing a layer to place skyrmion island, this system allows to utilize each island as one qubit. In another manner, by applying electric field and magnetic field selectively, encoding quantum information in skyrmion helicity of each island would be also possible.\cite{PhysRevLett.127.067201, PhysRevLett.130.106701}.


\begin{acknowledgments}
\noindent	
{\em Acknowledgments.---}
This research was supported by National Research Foudation Grant (2021R1A2C109306013).

\end{acknowledgments}

\appendix

\bibliography{biblio}


\end{document}